\newcommand*{\wn}{cm$^{-1}$}
\newcommand*{\AX}{A$^1\Pi$ - X$^1\Sigma^+$}

\newcommand*{\As}{A$^1\Pi$}

\documentclass[aip,jcp]{revtex4-1}
\usepackage{graphics}
\usepackage{dcolumn}
\usepackage[version=3]{mhchem}
\usepackage[english]{babel}
\usepackage{dcolumn}
\usepackage{graphicx}
\usepackage{amsmath}
\usepackage{tabularx}
\usepackage{bigstrut}
\usepackage{longtable}
\usepackage{rotating}
\usepackage{comment}
\usepackage{float}

\begin{document}

\title{High-precision laser spectroscopy of the CO \AX (2,0), (3,0) and (4,0) bands
}

\author{M. L. Niu}
\affiliation{Department of Physics and Astronomy, and LaserLaB, VU University, De Boelelaan 1081, 1081 HV Amsterdam, The Netherlands}
\author{F. Ramirez}
\affiliation{Department of Physics and Astronomy, and LaserLaB, VU University, De Boelelaan 1081, 1081 HV Amsterdam, The Netherlands}
\affiliation{Department of Physics, University of San Carlos, Cebu City 6000, Philippines}
\author{E. J. Salumbides}
\affiliation{Department of Physics and Astronomy, and LaserLaB, VU University, De Boelelaan 1081, 1081 HV Amsterdam, The Netherlands}
\affiliation{Department of Physics, University of San Carlos, Cebu City 6000, Philippines}
\author{W. Ubachs}
\affiliation{Department of Physics and Astronomy, and LaserLaB, VU University, De Boelelaan 1081, 1081 HV Amsterdam, The Netherlands}

\date{\today}

\begin{abstract}
\label{abstract}
High-precision two-photon Doppler-free frequency measurements have been performed on the CO \AX\ fourth-positive system (2,0), (3,0), and (4,0) bands. Absolute frequencies of forty-three transitions, for rotational quantum numbers up to $J = 5$, have been determined at an accuracy of $1.6\times10^{-3}$ \wn, using advanced techniques of two-color 2+1' resonance-enhanced multi-photon ionization, Sagnac interferometry, frequency-chirp analysis on the laser pulses, and correction for AC-Stark shifts. The accurate transition frequencies of the CO \AX\ system are of relevance for comparison with astronomical data in the search for possible drifts of fundamental constants in the early universe. The present accuracies in laboratory wavelengths of $\Delta\lambda/\lambda = 2 \times 10^{-8}$ may be considered exact for the purpose of such comparisons.

\end{abstract}

\maketitle

\section{Introduction}

The carbon monoxide molecule is one of the most thoroughly studied molecules in spectroscopy, from the microwave and infrared to the optical and vacuum ultraviolet parts of the electromagnetic spectrum. This second most abundant molecule in the Universe plays an important role in the chemical dynamics of the interstellar medium, as well as in Earth-based combustion, atmospheric, and plasma science. The \AX\ system of CO, known as the fourth positive system, connects the electronic ground state of the molecule to its first excited state of singlet character, and it signifies the onset of strong dipole-allowed absorption features in the vacuum ultraviolet range. The spectroscopy of the \AX\ bands has, since the first pioneering study by Deslandres in 1888,~\cite{Deslandres1888} been investigated by many authors (notably~\cite{Simmons1969,Lefloch1985}) culminating in a detailed analysis of the \As\ state and the many perturbing states by Field~\emph{et al.}.~\cite{Field1972}

The present precision-frequency investigation of the (2,0), (3,0) and (4,0) bands of the \AX\ system, by two-photon Doppler-free laser spectroscopy, is part of a program to apply the vacuum ultraviolet absorption lines of CO in a search for a variation of the proton-electron mass ratio at high redshift.~\cite{Salumbides2012} For this purpose the low-$J$ transitions in the CO absorption bands should be calibrated at an accuracy better than $\Delta\lambda/\lambda < 10^{-7}$, and the perturbations with other electronic states should be analyzed as accurately as possible, to obtain reliable line intensities. Highly accurate transition frequencies of lines in \AX\ (0,0) and (1,0) bands were reported,~\cite{Salumbides2012} as well as a detailed analysis of perturbations in the \As, $v=0$ and $v=1$ levels.~\cite{Niu2013}
Previously, the \AX (2,0), (3,0) and (4,0) bands have been investigated, at an accuracy of 0.06 \wn, by Lefloch, although the results were only partially published~\cite{Lefloch-thesis} in semi-open literature. In their compilation of CO-data for use in astronomy Morton and Noreau~\cite{Morton1994} have included these data of Le Floch. Here the laser-based methods of Ref.~\cite{Salumbides2012}, providing improved accuracy, are extended to the (2,0), (3,0) and (4,0) bands.

\section{Experimental details}

\begin{figure*}
\resizebox{1\textwidth}{!}{\includegraphics{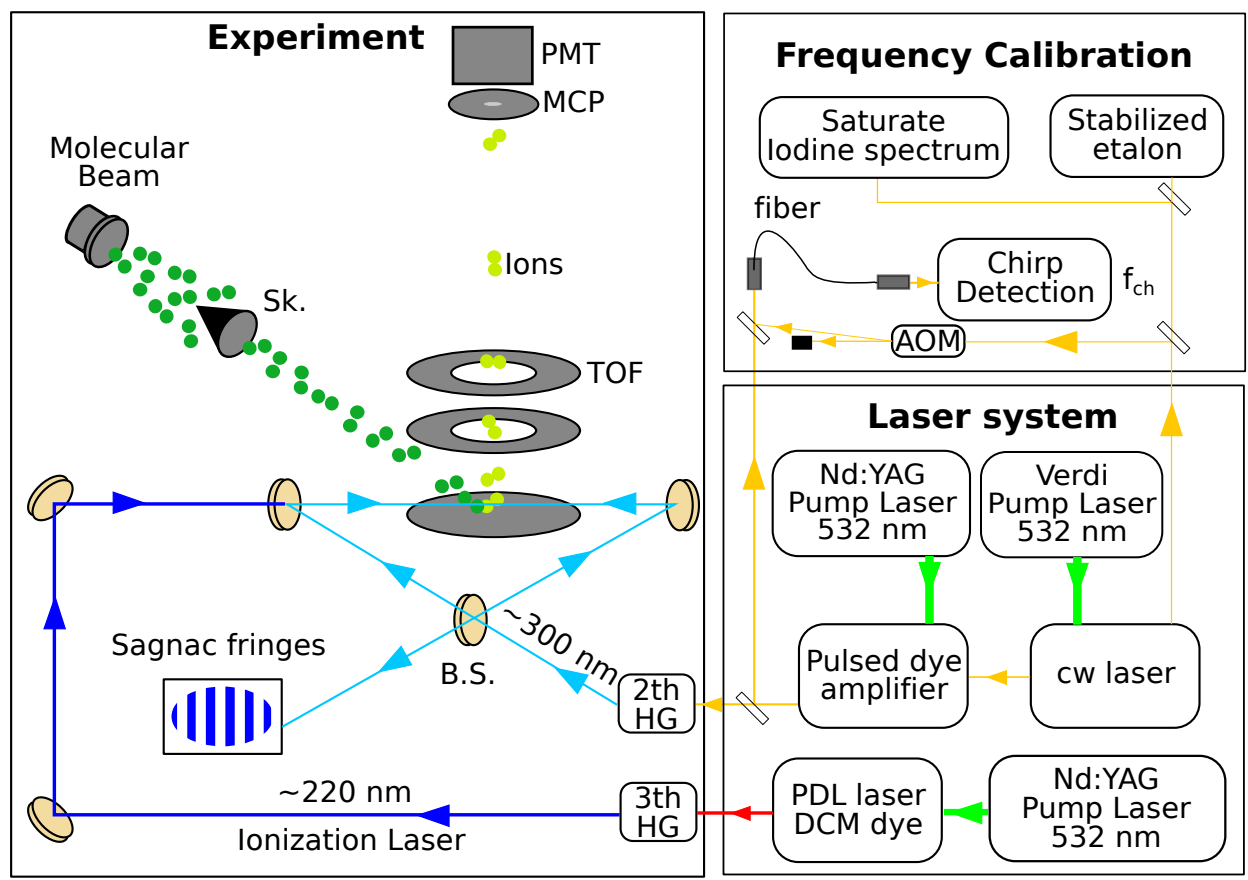}}
\caption{(Color online) A schematic layout of the experimental setup, with the laser system, the frequency calibration setup and the molecular beam apparatus with the counter-propagating laser beams in a Sagnac configuration for Doppler-free two-photon spectroscopy. B.S.: beam splitter; Sk.: skimmer; MCP: microchannel plate; PMT: photomultiplier tube; TOF: time of fight. See text for further details.}
\label{setup}
\end{figure*}

In this study, a narrowband pulsed-dye-amplifier (PDA) laser system is used in a two-photon Doppler-free excitation scheme for a high-resolution spectroscopic investigation of the CO \AX\ (2,0), (3,0) and (4,0) bands. The measurement setup is schematically shown in Fig.~\ref{setup}.
The PDA-laser system employed has been described previously~\cite{Eikema1997} as well as its application to high-resolution studies in CO.~\cite{Salumbides2012,Ubachs1997} The narrowband PDA system is built as a three-stage traveling-wave amplifier, seeded by a Coherent-899 continuous wave (cw) ring dye laser, and pumped by a pulsed Nd:YAG laser; it delivers nearly Fourier-Transform limited laser pulses of 5 ns duration. The repetition rate of the pulsed laser is 10 Hz. For the present investigation, the cw ring-dye laser was routinely operated in the range 560-590 nm, while running on Rhodamine-6G, but in the PDA system two different dyes were used to cover the three different bands: Rhodamine-B ($580-590$ nm) for the \AX\ (2,0) and (3,0) bands, and Rhodamine-6G  ($560-580$ nm) covering the \AX\ (4,0) band.

The output of the PDA laser passes through a KDP (Potassium Dihydrogen Phosphate) crystal for frequency doubling. Thereupon the laser beam is split in two on a beamsplitter and aligned in the geometry of a Sagnac interferometer to produce exactly counter-propagating beams for avoiding residual Doppler-shifts.~\cite{Hannemann2007} To systematically assess AC-Stark effects, different focusing conditions are employed.
An auxiliary laser beam at 220 nm is implemented for ionization of the A$^1\Pi$ excited state population. By this means the 2+1' resonance-enhanced multi-photon ionization (REMPI) scheme efficiently produces signal. In order to reduce AC-Stark effects by the ionization laser, its pulse timing is delayed by 10 ns with respect to the narrowband spectroscopy laser.

The CO molecular beam is produced by a pulsed solenoid valve (General Valve, Series 9) in a source chamber, before entering an interaction chamber via a skimmer (1.5 mm diameter). The collimated CO molecular beam is interrogated by the laser beams under collision-free conditions. CO ions produced in the 2+1' REMPI process are accelerated via pulsed extraction voltages applied after the laser pulses, to avoid DC-Stark effects. The ions are subsequently mass selected in a time-of-flight tube before detection on a multichannel plate (MCP).

For calibrating the frequency of the transitions on an absolute scale, saturated I$_{2}$ absorption spectroscopy is used, resolving hyperfine components, which are known to 1 MHz accuracy.~\cite{Velchev1998} For frequency interpolation between CO and I$_2$ resonances the transmission peaks of an etalon length stabilized to a He-Ne laser (with FSR = 150.01$\pm$0.01 MHz), were recorded simultaneously.

An important improvement over the previous laser-based study investigating the \AX\ (0,0) and (1,0) bands,~\cite{Salumbides2012} is that a frequency chirp-analysis setup has been implemented, following methods presented in Refs.~\cite{Fee1992,Eikema1997}, thus achieving higher precision.
In the present setup an acousto-optic modulator (AOM - Gooch \& Housego) shifts the frequency of the cw laser by 595 MHz.  The beatnote of this shifted cw laser frequency and the pulsed output of the PDA is detected by a fast photodiode, after the wave fronts are spatially overlapped by propagating through a single-mode fiber. The beat-note signal is then detected on a fast oscilloscope (Tektronix TDS7404) with 1 GHz analog bandwidth and 4 Gs/s sampling. The instantaneous frequency over the duration of the pulse is then determined as well as the frequency chirp over the pulse duration. The resulting frequency determinations of the CO resonances are finally corrected for this chirp effect.

\section{Results and discussion}

\begin{figure}
\resizebox{1\textwidth}{!}{\includegraphics{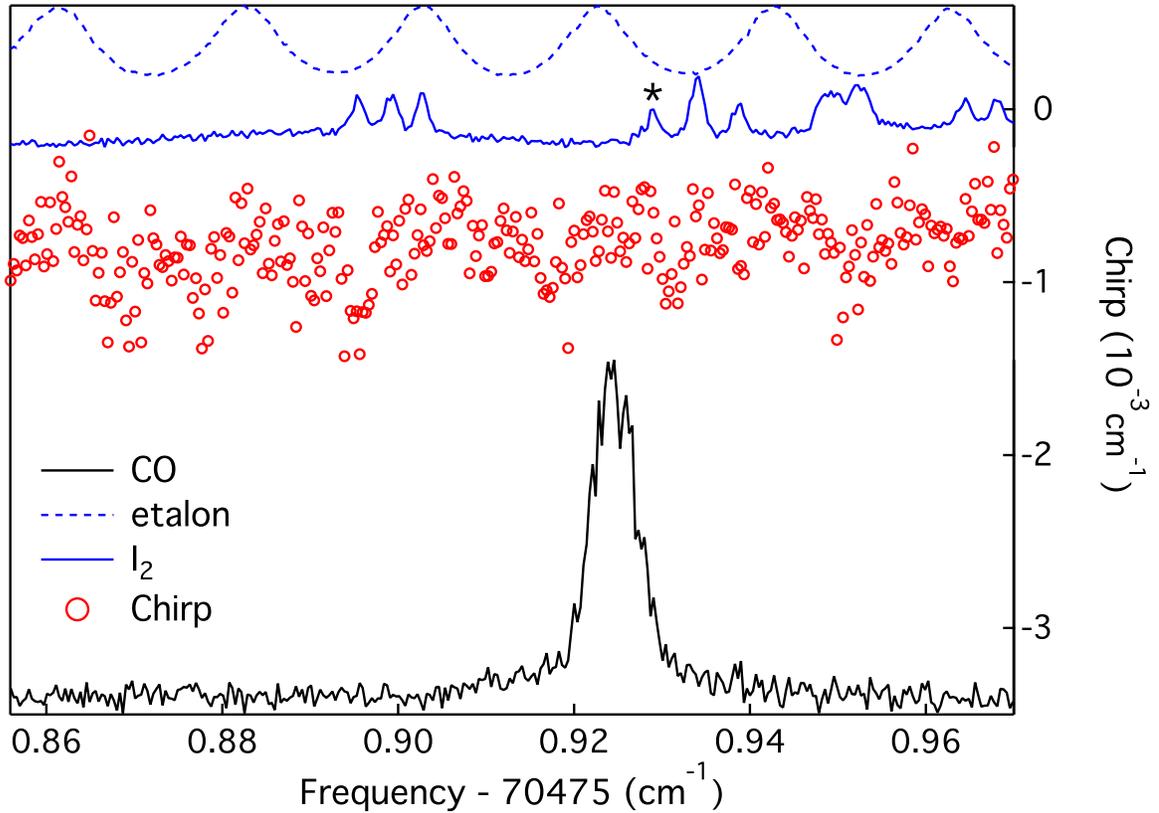}}
\caption{Typical recording of the S(0) transition in the CO \AX (4,0) band, as measured via two-photon laser excitation, plotted as the (black) solid line. The (blue) dashed line and the (blue) full line represent etalon markers and the saturated iodine spectrum used for interpolation and calibration. The (*) indicates the a4 (7,7) hyperfine component of the B-X (25,3) R(69) iodine line, which was used for the calibration. The (red) circles represent the on-line measured values for the chirp-induced frequency offset at the transition frequency.}
\label{PDAspec}
\end{figure}

In the present study two-photon resonances in the \AX\ (2,0), (3,0) and (4,0) bands of CO are measured by two-photon laser spectroscopy.
As an example, Fig.~\ref{PDAspec} shows a typical recording of the S(0) line in the \AX\ (4,0) band, plotted with simultaneous recordings of etalon markers, the saturated iodine spectrum for absolute calibration, and results of the on-line chirp analyses for each of the data points. The full width at half maximum (FWHM) of the CO transitions are around 200 MHz, limited by the bandwidth of the PDA laser.

Values for the transition frequencies were obtained from the calibration procedure, with corrections for the chirp and AC-Stark analysis, and multiplying by a factor of four for frequency-doubling and two-photon excitation.
Resulting two-photon transition frequencies are listed in Table~\ref{PDA-Lines}. In total, 43 lines in the \AX\ (2,0), (3,0) and (4,0) bands have been measured with an absolute accuracy of 0.0016 \wn, corresponding to 50 MHz, or a relative accuracy of $\Delta\lambda/\lambda = 2 \times 10^{-8}$. In view of the fact that excitation takes place in a molecular beam, only transitions originating in low rotational quantum states ($J<6$) could be observed. Lines are denoted with the usual O($J$) to S($J$) branches for transitions with $\Delta J=-2$ to $+2$.
It is noted that the two-photon R(0) transition is forbidden.~\cite{Bonin1984}

\begin{table}
\begin{center}
\begin{small}
\caption{Two-photon transition frequencies in the CO $A^1\Pi$ - $X^1\Sigma^+$ (2,0), (3,0) and (4,0) bands as obtained in the present study. Values in \wn.}
\label{PDA-Lines}
\begin{tabular}
{l@{\hspace{20pt}}c@{\hspace{20pt}}c@{\hspace{20pt}}c}
\colrule
\colrule
\multicolumn{1}{c}{Line} & \multicolumn{1}{c}{(2,0)} & \multicolumn{1}{c}{(3,0)} & \multicolumn{1}{c}{(4,0)}  \\
\colrule
S(0)	&	67\,685.126	&	69\,097.776	&	70\,475.924	\\
S(1)	&	67\,690.629	&	69\,103.134	&	70\,481.077	\\
S(2)	&	67\,695.397	&	69\,107.716	&	70\,485.399	\\
S(3)	&	67\,699.437	&	69\,111.519	&		\\
S(4)	&	67\,702.744	&	69\,114.546	&		\\
R(1)	&	67\,681.278	&	69\,093.930	&	70\,472.126	\\
R(2)	&	67\,682.932	&	69\,095.443	&	70\,473.460	\\
R(3)	&	67\,683.852	&	69\,096.179	&	70\,473.954	\\
R(4)	&	67\,684.038	&	69\,096.137	&		\\
R(5)	&		&	69\,095.318	&		\\
Q(1)	&	67\,675.051	&	69\,087.795	&	70\,466.088	\\
Q(2)	&	67\,673.592	&	69\,086.240	&	70\,464.388	\\
Q(3)	&	67\,671.404	&	69\,083.910	&	70\,461.854	\\
P(2)	&	67\,667.360	&	69\,080.104	&	70\,458.417	\\
P(3)	&		&	69\,074.706	&	70\,452.902	\\
P(4)	&		&	69\,068.531	&		\\
O(3)	&	67\,655.826	&	69\,068.571	&	70\,446.863	\\
\colrule
\colrule
\end{tabular}
\end{small}
\end{center}
\end{table}

The contributions to the overall uncertainty in the measurement accuracy are determined by statistical effects, AC-stark shifts, residual Doppler shift, and a chirped-induced frequency offset. In the following the two major sources of uncertainty will be discussed one by one. The contribution by the Doppler shift and the etalon non-linearity have been discussed previously.\cite{Salumbides2012}

\subsection{AC-Stark effect}

\begin{figure}
\resizebox{1\textwidth}{!}{\includegraphics{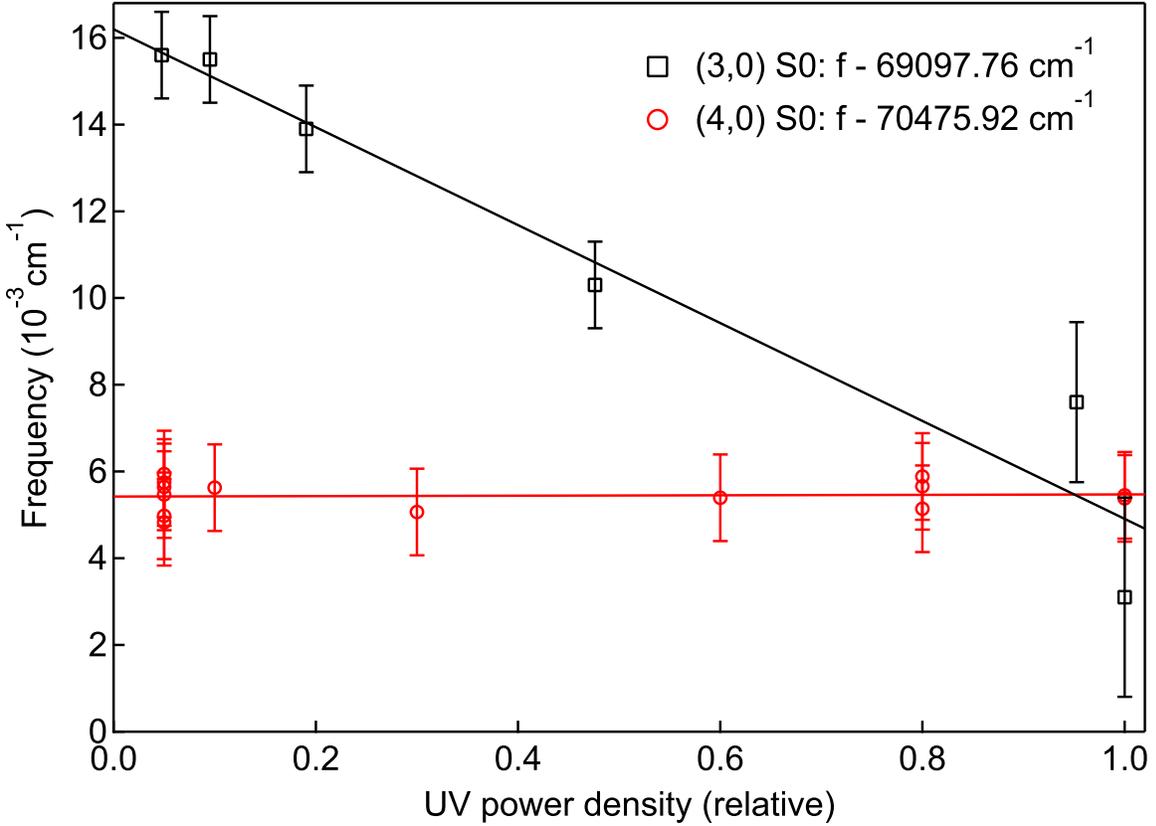}}
\caption{Assessment of the AC-Stark effect on the two-photon S(0) lines in the \AX\ (4,0) and (3,0) bands of CO. The (red) circles refer to transition S(0) in the (4,0) band, and the (black) squares to transition S(0) in the (3,0) band. The solid lines represent linear fits for extrapolation to zero intensity.}
\label{AC-stark}
\end{figure}

AC-Stark frequency shifts can be induced by the power density of the probe laser radiation. In order to reduce a possible AC-Stark shift, a separate ionization laser was used, which was delayed by 10 ns to avoid time overlap with the spectroscopy step. The shift induced by the spectroscopy laser can be analyzed and corrected for by performing an extrapolation to zero power density. Figure~\ref{AC-stark} shows the AC-Stark power extrapolation for the two-photon S(0) lines in the \AX\ (3,0) and (4,0) bands of CO.

From a detailed analysis it follows that the two-photon lines in the  (2,0), (3,0) and (4,0) bands exhibit different slopes in the AC-stark plots. For lines in the (2,0) and (4,0) bands, the AC-Stark shift appears to be smaller than the statistical error, as shown in the example of Fig.~\ref{AC-stark}. However, for the lines in the (3,0) band the AC-Stark effect is found to be much larger as exemplified for the S(0) transition. Such differences in slopes were also observed in the case of (0,0) and (1,0) bands.~\cite{Niu2013} All the measured transitions in the (3,0) band were independently extrapolated to zero power, and corrected for the AC-Stark shift. The AC-Stark effects are a measure for the polarizability of the quantum states involved, and the differences between bands indicate that the \As, $v=3$ levels have a strong polarizability. Further analysis of this phenomenon is beyond the scope of the present paper.

\subsection{Frequency chirp in the PDA}
In the frequency calibration procedure, the absolute frequency of the cw dye laser is determined from the saturated I$_{2}$ absorption spectra providing reference points known to 1 MHz accuracy.~\cite{Velchev1998} Since the CO spectra were recorded with the UV-upconverted pulsed output of the PDA this requires an assessment of the frequency offset between the pulsed output of the pulsed dye amplifier (PDA) and the cw laser seeding the PDA, which is referred to as chirp. The physical mechanisms of frequency chirp in dye amplifiers, associated with time-dependent index of refraction and time-dependent gain phenomena, have been well documented.~\cite{Fee1992,Melikechi1994}

The cw-pulse offset frequency is measured by heterodyning the amplified pulsed output of the PDA system with part of the cw-seed that is shifted by 595 MHz using an AOM, as shown in Fig.~\ref{setup}. The resulting beat signal measured on a fast photo-detector is recorded with a 4 Gs/s sampling oscilloscope for further analysis. All laser beams are combined and propagated through single-mode optical fibers in order to overlap wave fronts and stabilize the alignment of the chirp measurement. We follow procedures of chirp signal analysis involving frequency filtering and Fourier-transformation as described in Refs.~\cite{Hannemann2006, Fee1992, Eikema1997}.

\begin{figure}
\resizebox{1\textwidth}{!}{\includegraphics{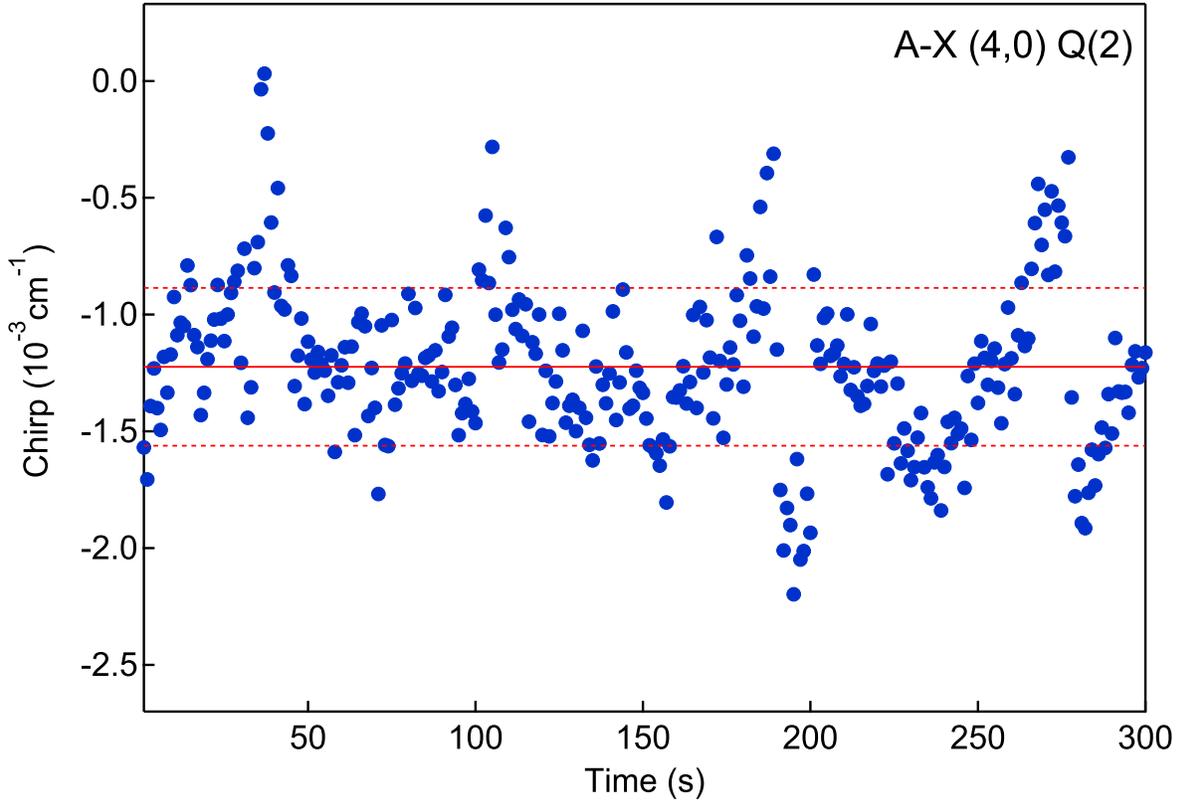}}
\caption{Deduced offset between cw and PDA laser frequency as a result of chirp over a 300 s time interval with PDA laser parked on the Q(2) line in the CO \AX\ (4,0) band. The (red) solid line indicates the averaged offset value of 37 MHz at the two-photon transition. The (red) dashed lines indicate the ($\pm\sigma$) standard deviation.}
\label{chirp_statistics}
\end{figure}

\begin{figure}
\resizebox{1\textwidth}{!}{\includegraphics{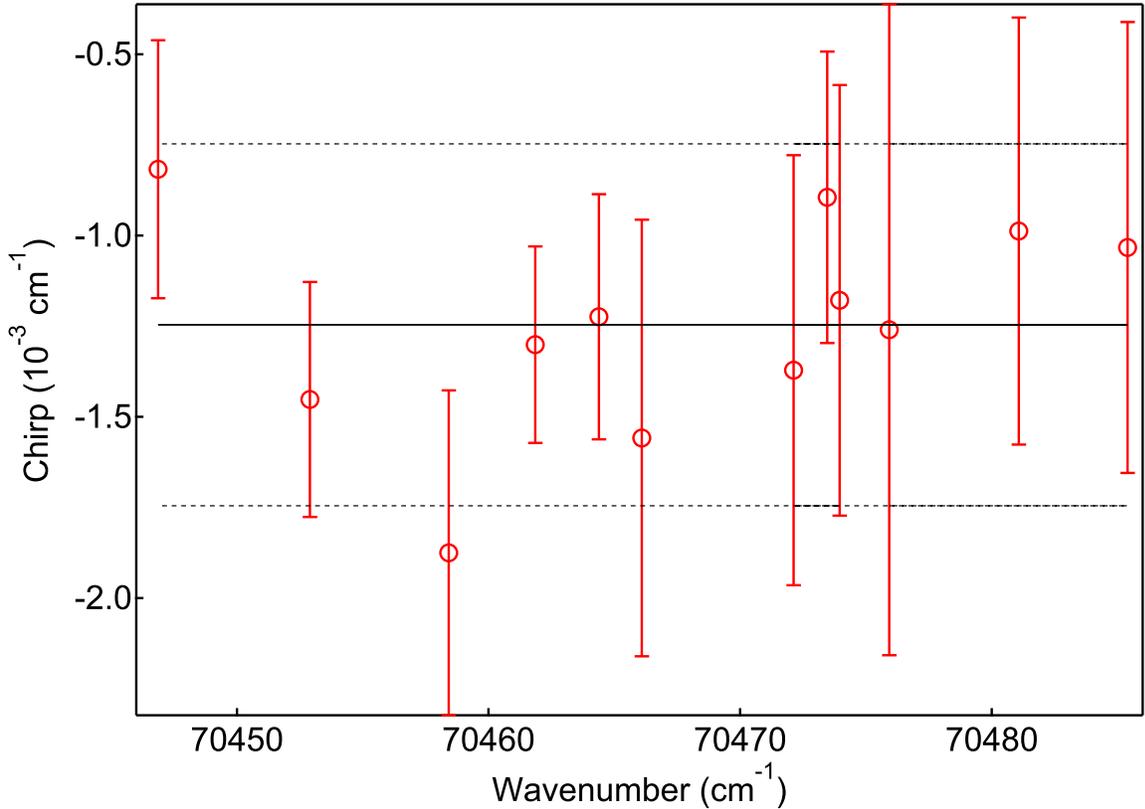}}
\caption{The chirp shift for different transitions in the \AX (4,0) band of CO. The (black) solid line is the averaged chirp shift in the frequency range of the \AX\ (4,0) band. The dashed lines represent the ($\pm\sigma$) standard deviation for the offsets.}
\label{chirpfrequency}
\end{figure}

Figure~\ref{chirp_statistics} shows a chirp-offset measurement recorded at the frequency of the Q(2) line in the CO \AX\ (4,0) band. The chirp-induced shift is measured over a duration of 300 s at 1 point/s with 10 times average. The average value of the offset is 37 MHz at the two-photon transition frequency, with a standard deviation of 10 MHz. The oscillatory structure in Fig.~\ref{chirp_statistics} is found to be due to a changing polarization during transmission of the cw laser in the fiber. The reduction in beat intensity then leads to an inaccurate determination of the frequency.
Such averaged cw-PDA offset was measured during recordings of the different transitions in CO \AX\ (4,0) band, results of which are shown in Fig.~\ref{chirpfrequency}. These measurements demonstrate that the chirp-induced offset values do not change over the small frequency ranges of CO resonance lines in a single band, so that averaged values are used for correcting the final transition frequencies.


The frequency chirp is known to vary over the spatial wave front of the laser beam.~\cite{Hannemann2006} This phenomenon was also systematically investigated and a maximum variation in the chirp-offset of 24 MHz was found, which is included as a statistical uncertainty for the chirp-offset in the error budget.

\begin{table}
\caption{\label{PDA-uncertainty}
Estimated contributions to the uncertainty in the two-photon transition frequencies for the \AX\ bands of CO.
}
\begin{tabular}{lc@{\hspace{40pt}}r@{.}l}
\colrule
\colrule
\textrm{Source}&
\textrm{MHz}&
\multicolumn{2}{c}{\textrm{\wn}}\\
\colrule
PDA chirp		    &30 & 0&001  \\
AC-Stark shift		&20 & 0&000\,7  \\
Line-fitting		&2 & 0&000\,07  \\
I$_2$ calibration	&2 & 0&000\,07  \\
Etalon nonlinearity	&2 & 0&000\,07  \\
DC-Stark shift		&$<1$ & 0&000\,03  \\
Residual Doppler	&$<1$ & 0&000\,03  \\
Statistical	    	&30 & 0&001  \\
\colrule
Total			    &47 & 0&001\,6  \\
\colrule
\colrule
\end{tabular}
\end{table}

\subsection{Uncertainty estimates and consistency}
The error budget for the CO transition frequencies is summarized in Table~\ref{PDA-uncertainty}. The systematic corrections, associated with chirp and AC-Stark effect are applied to each transition. All transitions are recorded more than once, and on different days to confirm reproducibility of the results.
The entry for the statistical uncertainty (1$\sigma$ standard deviation) is derived from a weighted average over individual scans, and including the fitting uncertainty. The overall uncertainty in the absolute frequencies for the present data set of Table~\ref{PDA-Lines} results from taking individual errors in quadrature yielding $0.0016$ \wn.

\begin{figure}
\begin{center}
\resizebox{1\textwidth}{!}{\includegraphics{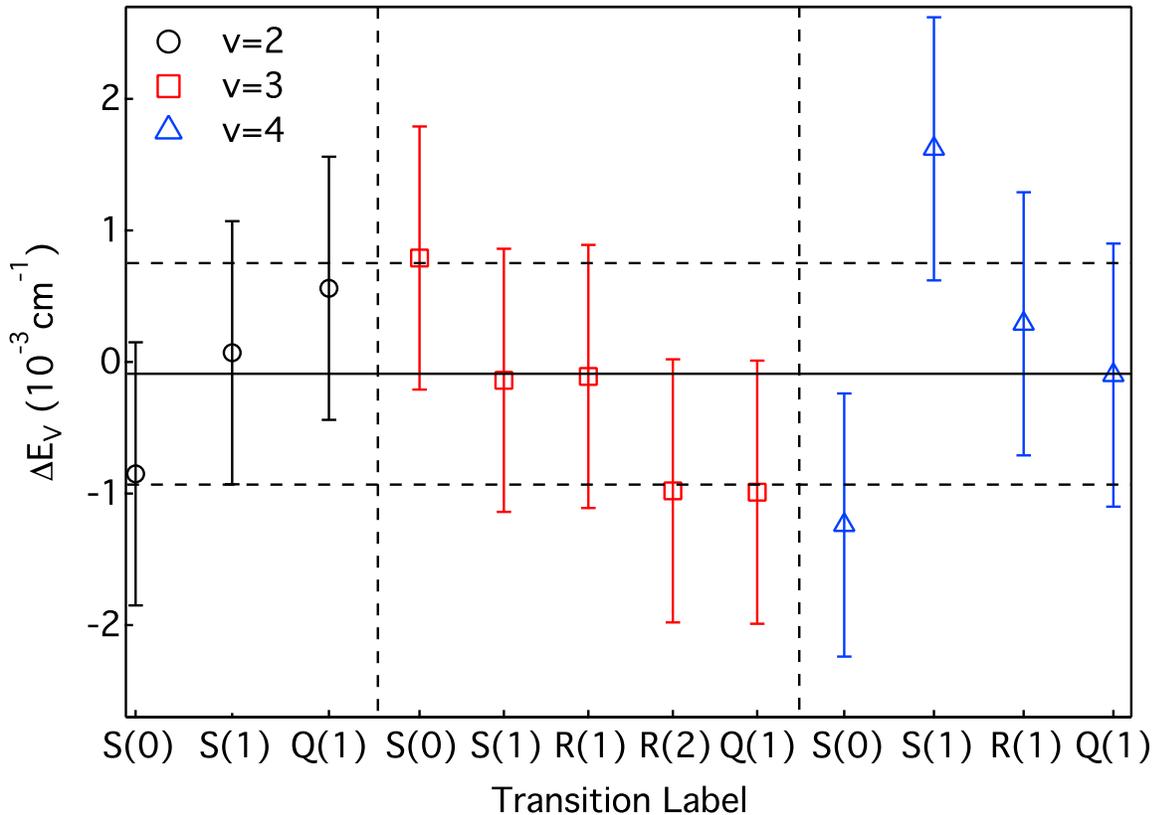}}
\caption{The combination differences in CO \AX\ (2,0), (3,0) and (4,0) bands, calculated and compared with the directly measured ground state splitting by Varberg and Evenson,~\cite{Varberg1992} represented as $\Delta E_{\rm{V}}$. The solid line is the averaged value $\Delta E_{\rm{V}}$ over all measured combination differences. The dashed lines indicate the ($\pm\sigma$) standard deviation for the comparison.}
\label{combination difference}
\end{center}
\end{figure}

The consistency of the error budget can be tested by calculating ground state combination differences from the experimental data. Combination differences between S($J$) and Q($J+2$), R($J$) and P($J+2$), and Q($J$) and O($J+2$) can be calculated and compared to the very accurately known ground state splittings from micro-wave and far-infrared rotational spectroscopy on the ground state of CO,~\cite{Varberg1992} yielding the difference $\Delta E_{\rm{V}}$. Such a comparison is graphically presented in Fig.~\ref{combination difference}, where the transitions used in the comparison are labeled. The comparison yields very good agreement within a standard deviation of only 0.0008 \wn. This is even smaller than the estimated measurement uncertainty and hence proves the internal consistency of the calibrations performed on the individual lines in the data set.

\section{Conclusion}
High-precision frequency measurements of 43 rotational lines in CO \AX\ (2,0), (3,0) and (4,0) bands (for $J<6$) have been performed in the collisionless environment of a molecular beam with an absolute accuracy of better than $2\times10^{-3}$ \wn, or a relative accuracy of $2 \times 10^{-8}$. 
The accuracy is an improvement over a previous analysis of the (0,0) and (1,0) bands, accurate to $3\times10^{-8}$,\cite{Salumbides2012,Niu2013} made possible by the chirp detection treatment.
The uncertainty is mainly determined by statistics, and by remaining effects of chirp on the laser pulses as well as an AC-Stark induced shift.
The accuracy obtained in this study represents a 10-fold improvement over our previous study.\cite{Salumbides2012}
After converting the accurate information on the level energies to one-photon vacuum ultraviolet transition frequencies these values can serve as calibration reference lines for synchrotron spectra,\cite{Niu2015} as well as implemented in comparisons with astrophysical spectra of high-redshift objects to derive a constraint on a possible variation of the proton-electron mass ratio on a cosmological time scale.~\cite{Salumbides2012}

\smallskip

This work was supported by Dutch Astrochemistry Program of NWO (CW-EW) and by a grant from the Templeton Foundation.

\end{document}